\documentstyle[epsf]{aipproc}

\begin{document}
\title{New Method of Extracting non-Gaussian Signals in the CMB}
\author{Jiun-Huei Proty Wu}
\address{Astronomy Department,
  University of California, Berkeley,
  CA 94720-3411, USA}

\maketitle

\begin{abstract}
Searching for and charactering the non-Gaussianity (NG) of a given field
has been a vital task in many fields of science,
because we expect 
the consequences of different physical processes to carry
different statistical properties.
  Here we propose a new general method of extracting non-Gaussian features
  in a given field,
  and then use simulated cosmic microwave background (CMB)
  as an example to demonstrate its power.
  In particular,
  we show its capability of detecting cosmic strings.
\end{abstract}


With the cosmological principle as the basic premise,
two currently competing theories 
for the origin of structure in the universe
are inflation \cite{Guth}
and topological defects \cite{VilShe,Wu1}.
Although 
the recent CMB observations seem to have favored the former \cite{maxiboom},
the latter can still coexist with it.
In particular,
the observational verification of defects will
have certain impact to the grand unified theory,
since they are an inevitable consequence of 
the spontaneous symmetry-breaking phase transition in the early universe.
In addition to the conventional study of the power spectra
of cosmological perturbations,
another way to distinguish these models is via
the search for intrinsic NG---while
the standard inflationary models predict Gaussianity,
theories like isocurvature inflation \cite{Pee}
and topological defects \cite{LSS-NG} generate NG.
Here we shall 
propose a new method of extracting the NG from a given field \cite{nmoens},
and then apply it to the CMB,
which is arguably the cleanest cosmic signals \cite{HuSugSil}.

The new method aims to nothing but
removing the `Gaussian' components: the mean and the power.
Using an $n$-dimensional field $\Delta({\bf x})$ as an example,
the method first
  Fourier transforms $\Delta({\bf x})$
  to yield $\widetilde{\Delta}({\bf k})$.
Then the power spectrum can be estimated as
  $C_k=\langle|\widetilde{\Delta}({\bf k})|^2\rangle_k/V^n$,
  where $V^n$ is the $n$-dimensional volume of the field
and $k\equiv |{\bf k}|$.
Next we define and calculate ($\forall {\bf k} \textrm{ with } C_k \neq 0$)
  \begin{equation}
  \label{Delta_p}
    \widetilde{\Delta}_P({\bf k})
    ={\left[
            \widetilde{\Delta}({\bf k})-
            \widetilde{\Delta}({\bf 0})\delta({\bf k})
           \right]}
     {{C_k^{-1/2}}}{P_k^{1/2}},
  \end{equation}
  where
  $\delta({\bf k})$ is a Dirac Delta and
  $P_k$ is a given function of $k$.
Finally,
$\widetilde{\Delta}_P({\bf k})$ is transformed back to the real space
  $  \Delta_P({\bf x})$.
Now the field $\Delta_P$ has a mean $\overline{\Delta}$ equal to zero 
and a power spectrum renormalized to $P_k$.
For the simplest case $P_k=1$,
the field $\Delta$ is `whitened' in the Fourier space,
and we shall use the superscript `W' to denote such whitened fields.
In the real space, this means
$
  \Delta = \Delta^{\rm W} \otimes D
     +\overline{\Delta}
$,
where $\otimes$ denotes a convolution,
and 
$
  D\equiv 
    \int dk^n
     {C_k^{1/2}} e^{i{\bf k\cdot x}}
$.
Thus 
 $\Delta$ is now decomposed into two parts:
the `Gaussian components', $\overline{\Delta}$ and $D$,
which carry the information in the mean and the power spectrum,
and the `NG component', $\Delta^{\rm W}$,
which possesses all the remaining information.
Therefore, if $\Delta$ is a Gaussian field,
then all samples in $\Delta^{\rm W}$ should appear uncorrelated
as pure white noise.
Otherwise $\Delta^{\rm W}$ would contain all the non-Gaussian features
\cite{nmoens}.
We note that the above treatment can be easily converted to the conventional
multipole transform for the CMB,
although we shall continue to use the Fourier convention,
which is appropriate for small CMB fields.
In this case, we have $\ell\equiv k$ and $C_\ell\equiv C_k$.
We also notice that 
the above new method is equivalent to the matrix manipulation
${\bf d}_P={\bf P}^{1/2} {\bf C}^{-1/2}{\bf d}$,
where ${\bf d}\equiv \Delta$, 
and ${\bf P}$ and ${\bf C}$ are the two-point correlation matrices
specified by $P_k$ (with $P_0=0$) and $C_k$ respectively.
This is similar but different from the Wiener filtering.

\begin{figure}[t]
  \centering
  \leavevmode\epsfxsize=2.6in\epsfbox{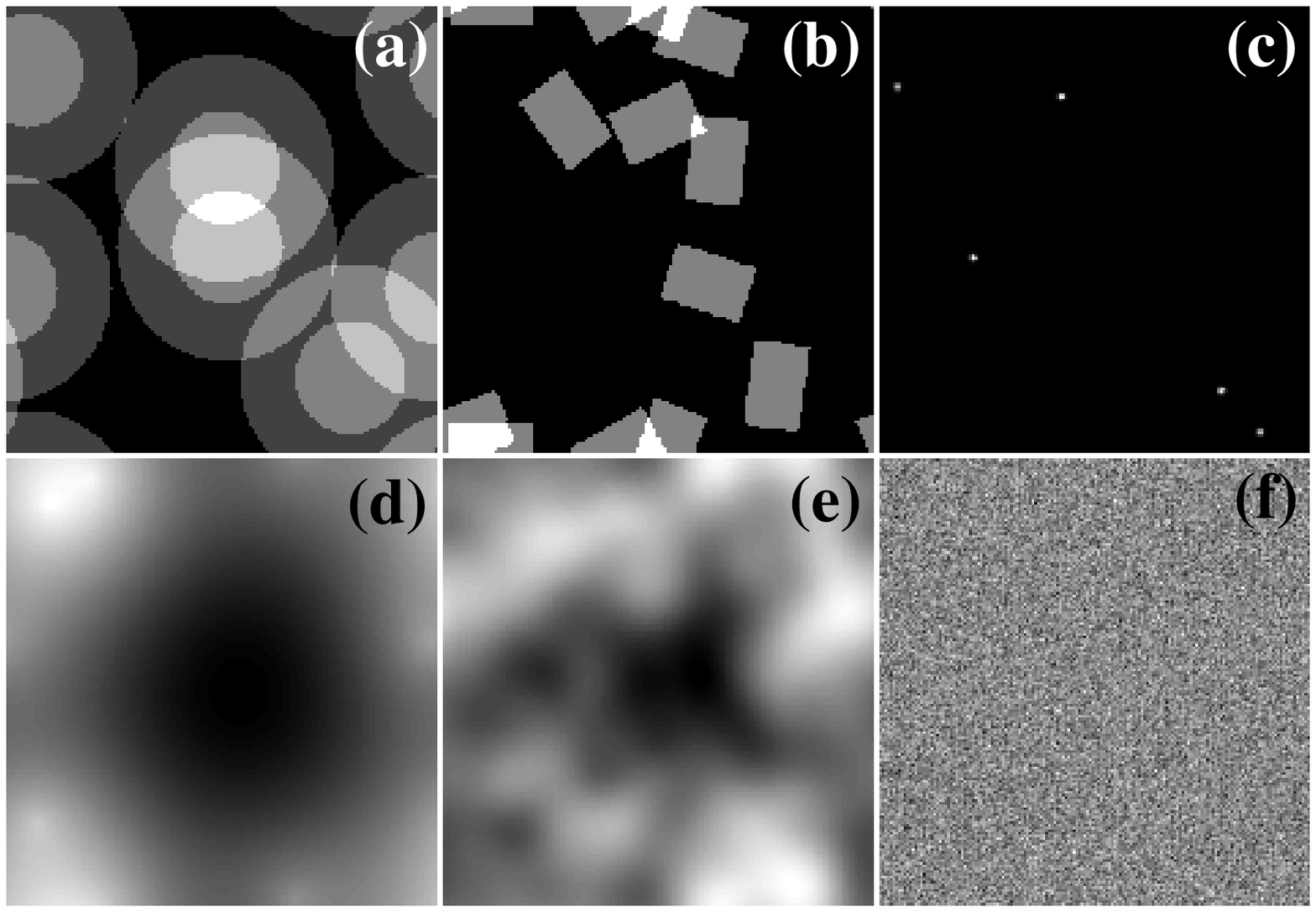}\hspace*{2mm}
  \leavevmode\epsfxsize=2.2in\epsfbox{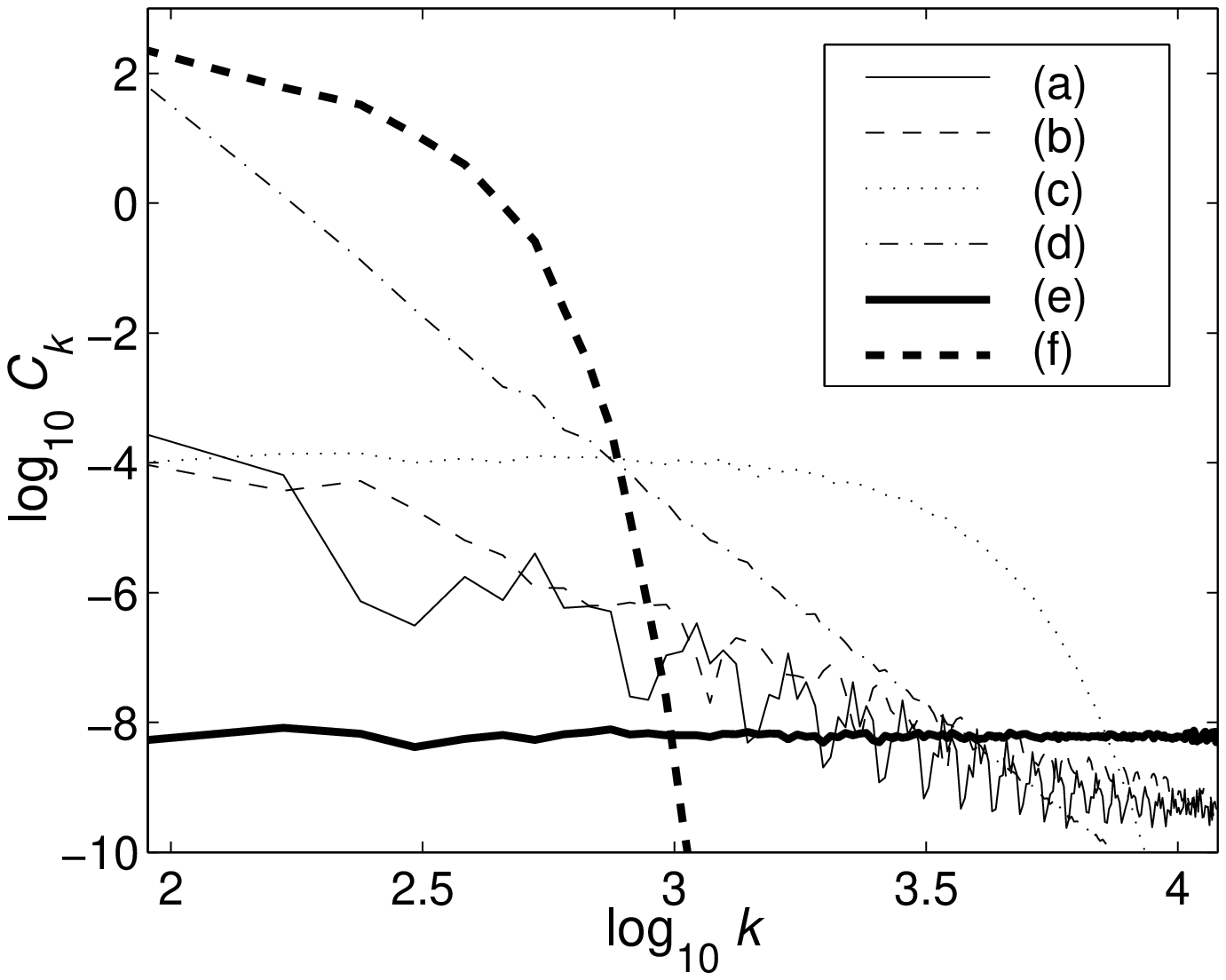}\\
  \caption
  {
Six different components in a simulated CMB map,
and their power spectra.
  }
  \label{fig-var-components}
\end{figure}
\begin{figure}[t]
  \centering
  \leavevmode\epsfxsize=2.5in \epsfbox{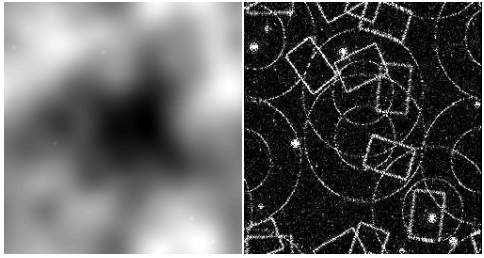}\\
  \caption
  {
A simulated CMB map $\Delta$ (left),
 and the extracted NG signal $(\Delta^{\rm W})^2$ (right).
    }
  \label{fig-test2}
\end{figure}

We now test this formalism using simulations.
Figure~\ref{fig-var-components}
shows six simulated CMB components:
(a)--(d) (where (d) contains 5 diffuse points) are non-Gaussian,
while (e) and (f) are Gaussian.
They are then linearly summed to yield
$\Delta=\sum_i \Delta_{(i)}$,
with  RMS ratios ((a)--(f))
$1:1:10:500:1000:0.2$ (Figure~\ref{fig-test2} left).
We then apply our method to obtain the extracted NG signal $\Delta^{\rm W}$
(Figure~\ref{fig-test2} right).

\begin{figure}[t]
  \centering
  \leavevmode\epsfxsize=2.9in \epsfbox{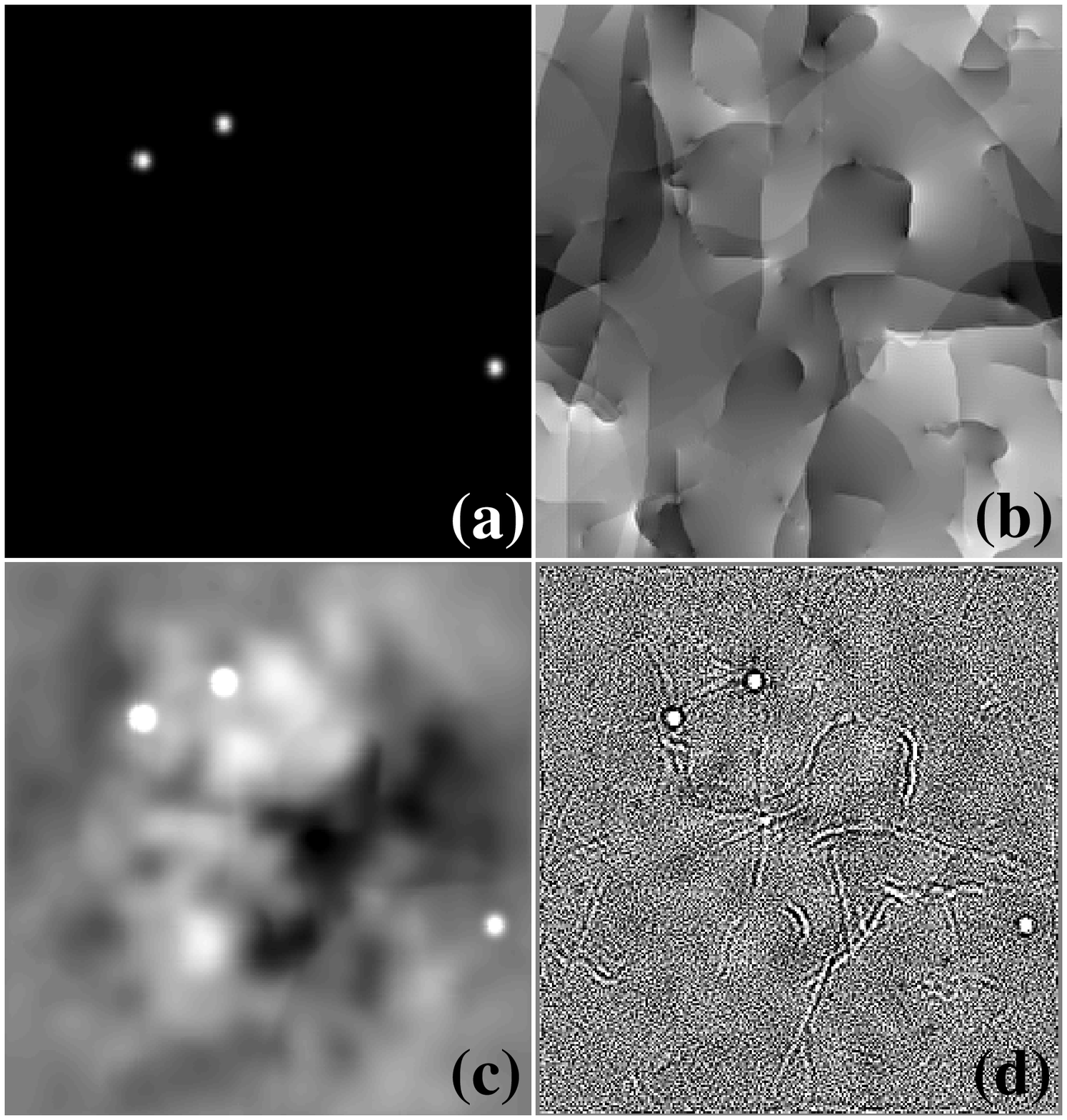} \hspace*{2mm}
  \leavevmode\epsfxsize=2.7in \epsfbox{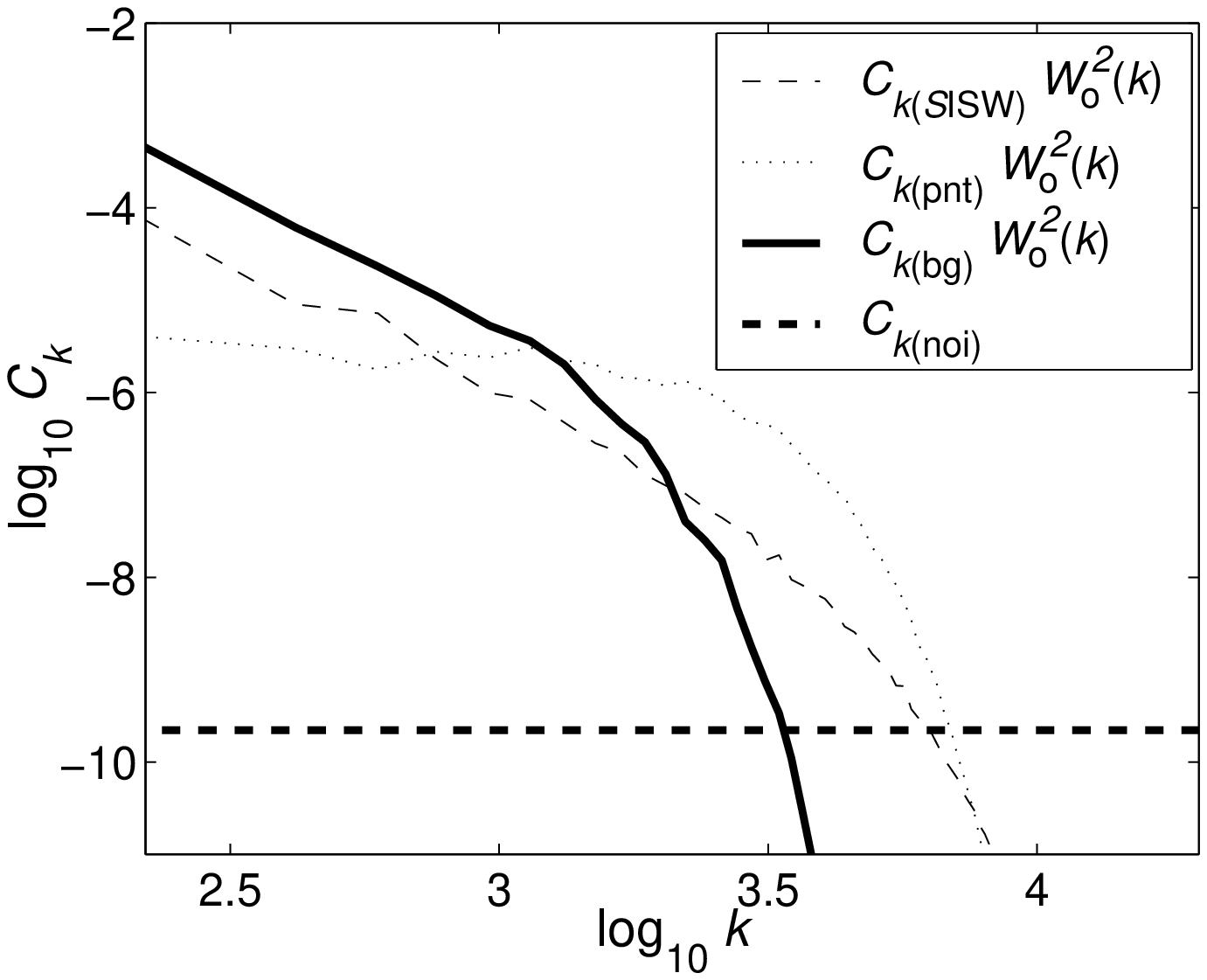}\\
  \caption
  []
  {
Simulated CMB, the extracted NG signal (d), 
and the power spectra (right).
    }
  \label{fig-str}
\end{figure}

In a second test,
we simulate a CMB field of $(2^\circ)^2$ (Figure \ref{fig-str}(c)):
$
  \Delta_{\rm s}
  =
  \left[
    \Delta W_{\rm p}
  \right]
  \otimes
  W_{\rm o}+\Delta_{\rm noi},
$
where
$\Delta=
\Delta_{\rm bg}$ (Gaussian background)
$+\Delta_{\cal S \rm ISW}$
(string-induced CMB\cite{Wu2001}; Figure \ref{fig-str}(b))
$+\Delta_{\rm pnt}$ (point source; Figure \ref{fig-str}(a))
with RMS ratios $5:1:2$,
$\Delta_{\rm noi}$ is a 5\% noise,
and 
$W_{i}$ with $i=$ `p' and `o' 
denote the primary and observing  beams respectively.
The  whitened field $\Delta_{\rm s}^{\rm W}$
is shown in Figure~\ref{fig-str} (d).

With even more tests,
the main observation remains the same:
in a field $\Delta=\Delta_{\rm (G)}+\Delta_{\rm (NG)}$,
regardless how stronger the $\Delta_{\rm (G)}$ is,
the NG features of $\Delta_{\rm (NG)}$ can always show up 
in the whitened field $\Delta^{\rm W}$
as long as
$C_{k{\rm (NG)}}$ dominates $C_{k{\rm (G)}}$
within a certain range of $k$.
In fact, 
this can be analytically proved \cite{nmoens}.
In addition,
the NG features of uncorrelated NG components do not mix up
in the extracted NG field $\Delta^{\rm W}$,
even if some of them dominate the others in power.

Finally we notice that
according to equation (\ref{Delta_p}),
in principle we can design a `window function' $P_k$
to keep the power only on scales where the NG components of a field dominate.
However,
in general we do not know what these scales are and
thus taking $P_k=1$ is optimal.
This may even enable us to find the NG signals of unknown physical processes.
We acknowledge the support from 
NSF KDI Grant (9872979) and
NASA LTSA Grant (NAG5-6552).


\end{document}